\documentclass{sig-alternate}
\usepackage{colortbl}
\usepackage{color}
\usepackage{url}
\usepackage{multirow}
\usepackage{arydshln}
\usepackage{pgf}
\usepackage{tikz}
\usepackage{pgfplots}
\usepackage{subfigure}
\begin{document}
\definecolor{light-gray}{gray}{0.80}
%

\title{Deep Learning Relevance: Creating Relevant Information (as Opposed to Retrieving it)}
%
%
%
%
%

\numberofauthors{2} 
%
\author{
%
%
\alignauthor
Christina Lioma\\
       \affaddr{Department of Computer Science}\\
       \affaddr{University of Copenhagen, Denmark}\\
       \email{c.lioma@di.ku.dk}
\alignauthor
Birger Larsen\\
       \affaddr{Department of Communication}\\
       \affaddr{Aalborg University Copenhagen, Denmark}\\
       \email{birger@hum.aau.dk}
\and
\alignauthor
Casper Petersen\\
       \affaddr{Department of Computer Science}\\
       \affaddr{University of Copenhagen, Denmark}\\
       \email{cazz@di.ku.dk}
\alignauthor
Jakob Grue Simonsen\\
       \affaddr{Department of Computer Science}\\
       \affaddr{University of Copenhagen, Denmark}\\
       \email{simonsen@di.ku.dk}
}
\maketitle
\begin{abstract}
What if Information Retrieval (IR) systems did not just retrieve relevant information that is stored in their indices, but could also ``understand'' it and synthesise it into a single document?
We present a preliminary study that makes a first step towards answering this question.

Given a query, we train a Recurrent Neural Network (RNN) on existing relevant information to that query. We then use the RNN to ``deep learn'' a single, synthetic, and we assume, relevant document for that query. We design a crowdsourcing experiment to assess how relevant the ``deep learned'' document is, compared to existing relevant documents. Users are shown a query and four wordclouds (of three existing relevant documents and our deep learned synthetic document). The synthetic document is ranked on average most relevant of all.

\end{abstract}



\section{Introduction}
\label{s:intro}
Deep neural networks aim to mimick the multiple layers of neurons that operate in the human brain in order to learn how to solve a wide range of interesting problems, like identifying photos \cite{KhoslaRTO15} or responding to web search queries \cite{SordoniBVLSN15,WuHLC15}. For Information Retrieval (IR), the main idea behind \textit{deep learning} is that, given enough input data (such as search engine queries or user clicks), a deep learning algorithm can learn the underlying semantics of this data (and hence emulate ``understanding'' it), and as a result lead to improved ways of responding to search queries. Google was recently reported \cite{Metz16} to having run a test that pitted a group of its human engineers against RankBrain, its deep learning algorithm. Both RankBrain and human engineers were asked to look at various web pages and decide which would rank highest by the Google search engine. While the human engineers were right 70\% of the time, RankBrain was right 80\% of the time \cite{Metz16}.

Motivated by these developments, we ask the following question: is it possible to train an IR system to learn and create \textit{new} relevant information, instead of just retrieving \textit{existing} indexed information? In other words, what if we could push IR systems one step further so that they do not just return the relevant information stored in their indices, but they synthesise it into a single, relevant document that, we assume, encompasses all indexed relevant information to that query?
We present a method for doing so in Section \ref{s:model}. Experiments with crowdsourced users (discussed in Section \ref{s:exp}) show that the new, ``deep learned'' synthetic documents (represented as word clouds) are considered on average more relevant to user queries, than the existing relevant indexed documents (also represented as wordclouds).


\section{Related Work}
\label{s:relw}
We briefly discuss applications of deep learning to IR. A more comprehensive overview of deep learning for web search in particular can be found in \cite{Li014}, chapter 7.2.5.  

A common application of deep learning to IR is for \textit{learning to rank}, covering a wide range of subtopics within this area, for instance from the earlier RankNet \cite{BurgesSRLDHH05}, and methods for hyperlinked webpages \cite{ScarselliYGHTM05}, to studies of listwise comparisons \cite{CaoQLTL07}, short text pairwise reranking \cite{Severyn:2015:LRS:2766462.2767738}, and elimination strategies \cite{TranPV16a}.

Deep learning has also been used to develop semantic or topic models in IR. Wang et al. \cite{WangMW07} ``deep learned'' both single and multiple term topics, which they integrated into a query likelihood language model for retrieval. 
Ranzato et al. \cite{RanzatoS08} used deep learning in a semi-supervised way to build document representations that retain term co-occurrence information (as opposed to only bag of words). 
Huang et al. \cite{HuangHGDAH13} and Shen et al. \cite{ShenHGDM14b} deep learned latent semantic models to map queries to their relevant documents using clickthrough data. 
Ye et al. \cite{YeQM15} also used clickthrough data to deep learn query - document relevance by modelling the proportion of people who clicked a document with respect to a query, among the people who viewed the document with respect to that query, as a binomial distribution. 
Lee et al. \cite{LeeACS15} used deep learning for fusing multiple result lists, while Liu et al. \cite{LiuGHDDW15} used deep neural networks for both web search ranking and query classification. 
Closest to our work, in terms of the deep learning methods used but not the problem formulation, is the work of Palanga et al. \cite{PalangiDSGHCSW15}, who used recurrent neural networks (RNNs) with Long Short-Term Memory (LSTM) cells (as we also do, discussed in Section \ref{s:model}) to create deep sentence embeddings, which they then used to rank documents with respect to queries. The similarity between the query and documents was measured by the distance between their corresponding sentence embedding vectors.  
Interesting is also the study of Mitra et al. \cite{MitraNCC16}, who used deep learning for reranking (with success) and ranking (with less success). Specifically, Mitra et al. used neural word embeddings, keeping both the input and the output projections. They then mapped the query words into the input space and the document words into the output space, and computed a query-document relevance score by aggregating the cosine similarities across all the query-document word pairs. They found this method effective for re-ranking top retrieved documents, but ineffective for ranking more candidate documents. 

Finally, the IR industry has not only adopted, but also opened up deep learning components to the public. Yahoo's CaffeOnSpark, Google's TensorFlow, and Microsoft's CNTK deep learning platforms are now open source, whereas Facebook has shared its AI hardware designs, and Baidu has unveiled its deep learning training software.

\section{Synthesising Relevant Information with Recurrent Neural Networks}
\label{s:model}
Given a query and a set of relevant documents to that query, we ask whether a new synthetic document can be created automatically, which aggregates all the relevant information of the set of relevant documents to that query. If so, then we reason that such a synthetic document can be more relevant to the query than any member of the set of relevant documents.   

We create such synthetic documents using RNNs. In order for the RNN model to learn a new synthetic document that does not previously exist, it needs to capture the underlying semantics of the query and its known relevant documents. We approximate this by feeding the text of the query and its known relevant documents as input to the RNN, using vector representations of characters\footnote{Such embeddings can also be created on a word level, but we do not experiment with this here. }, also known as embeddings. 

\begin{figure}
\centering
\scalebox{0.3}{
\includegraphics[]{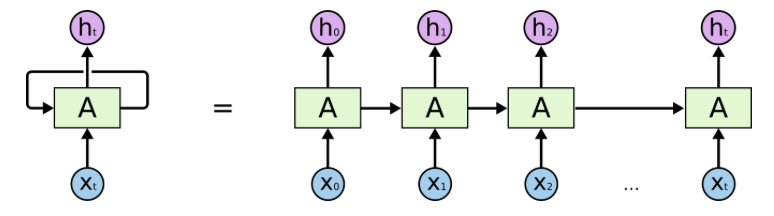}
}
\caption{\label{fig:RNNexample}An unrolled recurrent neural network. Example borrowed from Olah (2015) [13].}
\end{figure}

The order of characters is important for obtaining correct words. 
Given a character, the RNN takes as input its embedding and updates an internal vector (recurrent state) that functions as an order-sensitive summary of all the information seen up to that character (the first recurrent state is set to the zero vector). By doing so, the RNN parameterises a conditional probability distribution on the space of possible characters given the input encoding, and each recurrent state is used to estimate the probability of the next character in the sequence. The process continues until an end-of-document symbol is produced (see Figure \ref{fig:RNNexample} for an illustration). 

More specifically, the type of RNNs  we use are Long Short Term Memory (LSTM) networks \cite{HochreiterS97,Olah15}, which are capable of learning long-term dependencies, i.e. estimating the probability of the next character in the sequence based on the characters not just immediately preceding it, but also occurring further back.

In principle, any algorithm capable of generating a synthetic document from a set of input documents (and a query) can be used. 
We use RNNs due to their generally good performance in various tasks, and also to their ability to process unlimited-length sequences of characters.



%
%
%
%
%
%


%
%
\section{Evaluation}
\label{s:exp}
\subsection{Experimental Setup}

\begin{figure*}
\centering
\scalebox{0.4}{
\includegraphics[]{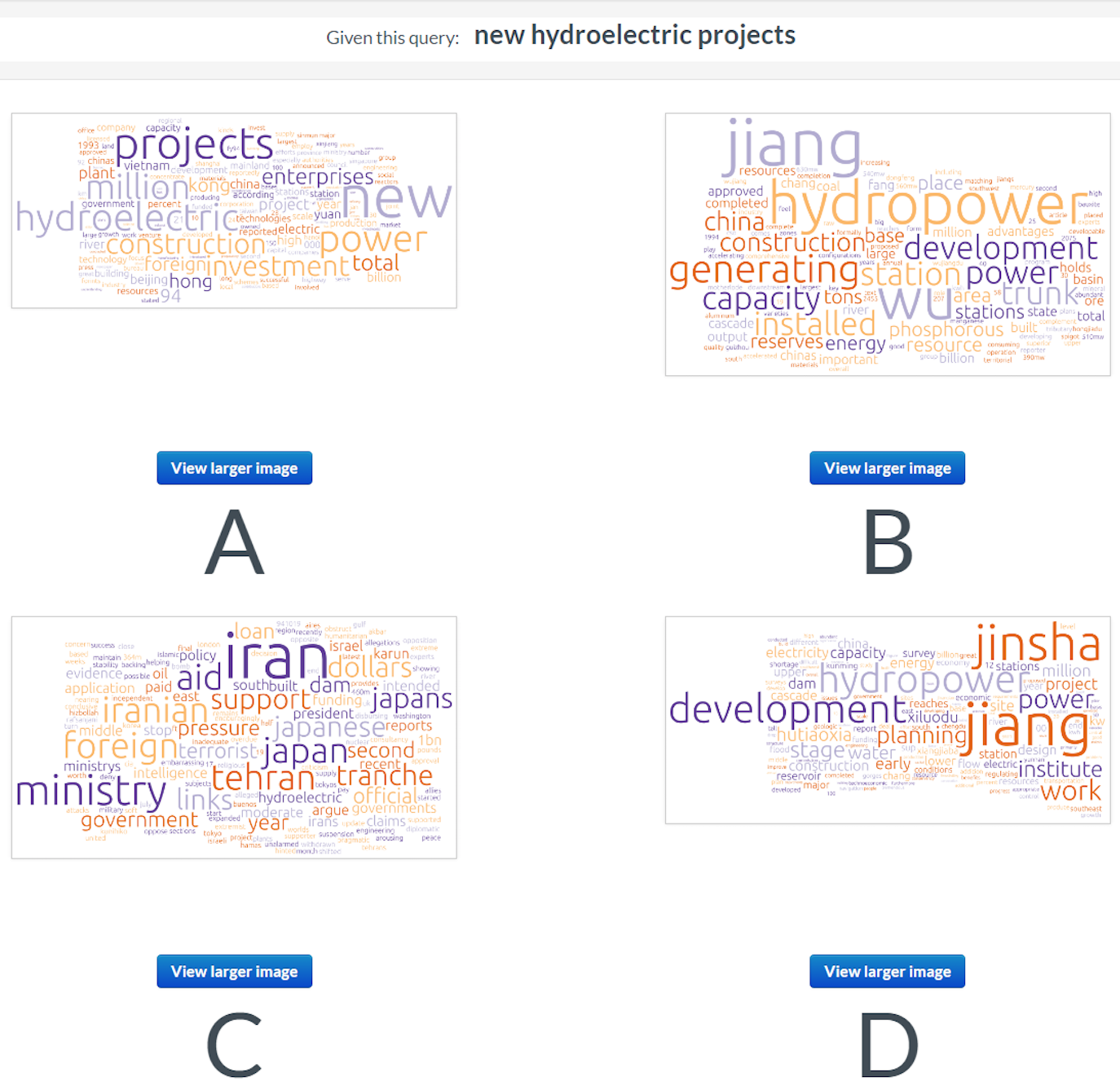}
}
\caption{\label{fig:example}Example query and wordclouds shown to Crowdflower users. In this example, the synthetic document's wordcloud is A. Out of a total of 10 users, 8 gave valid answers for this query. Wordcloud A was ranked on average at position 1.5 (5 users ranked it first, 2 users ranked it second, and 1 user ranked it third).}
\end{figure*}

\subsubsection{Experimental Design}
To assess to what extent our new deep learned synthetic document is relevant to the query, in comparison to any of the indexed known relevant documents, we carry out the following experiment. We use a TREC test collection, for which we have a list of previously assessed relevant documents to each query. For each query, we draw randomly three of its known relevant documents. Then, we present to the user the query text and also four documents: the three known relevant, and the deep learned synthetic document. For an easier and faster interpretation of the documents, we present these four documents as wordclouds, formed using only the 150 most frequent terms in each document. We present these four wordclouds not as a traditional ranked list, but in a 2 x 2 grid (see example in Figure \ref{fig:example}). The wordclouds are labelled A, B, C, D.  We control the order of presentation, so that the wordcloud of the synthetic document is rotated across all four positions an equal number of times. The task of the user is to rank these four wordclouds by their relevance to the query.

We experiment with the TREC Disks 4 \& 5 test collection (minus the Congressional Records) with title-only queries from TREC-6,7,8 (queries 301 - 450). 

\subsubsection{RNN training}
We train, for each query separately, a 3-layer, 512 neurons LSTM RNN, using the implementation of J. Johnson\footnote{\url{https://github.com/jcjohnson/torch-rnn}} on default settings. For each query, we concatenate the text of all its known relevant documents into a single character-based sequence,
using a new ``end-of-file'' character as file separator. We then repeatedly sample single characters from the model until an end-of-file
character appears in the sequence. More specifically, we do not use the whole text of the known relevant documents for a query; instead, we extract a context window of $\pm$ 30 terms around every query term in the document. The size of the context window is an ad-hoc choice aiming to trade-off data sparsity (if there are too few terms in the training set we cannot train the LSTM RNN) and noise (terms which may not contribute to the meaning of the query term). Using this setup, there was sufficient training data for 101 out of all 150 queries. 
Finally, we remove terms not found in the vocabulary of the collection and stop words.


\subsubsection{Crowdsourcing}
To assess the relevance of the wordclouds to the query, we crowdsourced users through CrowdFlower\footnote{\url{http://www.crowdflower.com}} (CF), defining four queries per job. 
Each query was assessed by 10 users. In addition to the example shown in Figure \ref{fig:example}, there were four boxes: (1) a box asking users to type their ranking of the four wordclouds, (2) a yes/no box asking users if they understand the query, (3) a comment box for optional feedback, and (4) a box asking users to type the two most salient terms of the wordcloud they ranked most relevant. We used this last box as a way to combat crowdsourcing misconduct.
We chose the highest quality CF users. There was a minimum time of 20 seconds specified per job. We did not use any restrictions on the crowdsourcing platforms that CF syndicates from, or on language, but we restricted participation to users from English-speaking and northern European countries\footnote{Participating users were from: USA, UK, Ireland, Canada, New Zealand, Sweden, the Netherlands.
}. Users were rewarded 0.1USD per job.

\begin{figure*}
\centering
\scalebox{0.4}{
\includegraphics[]{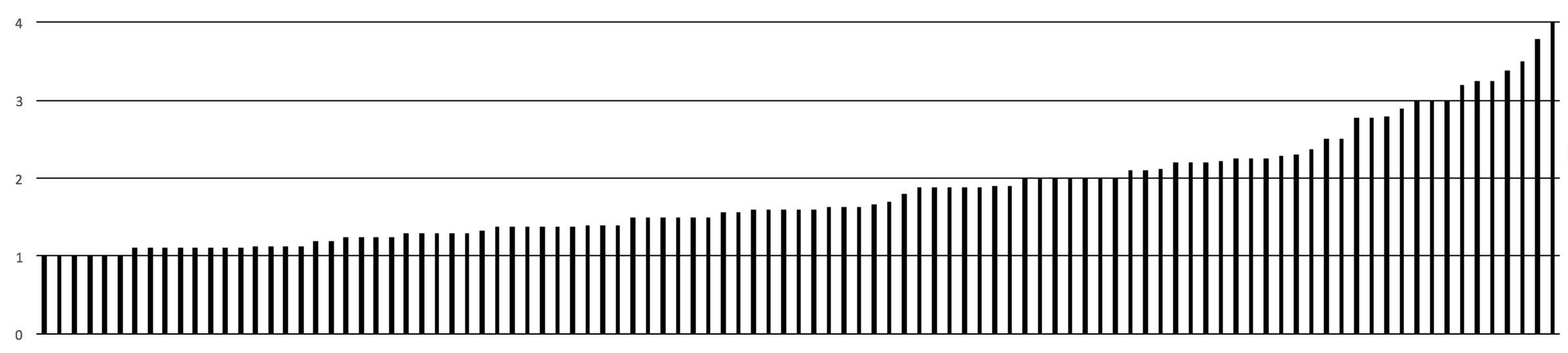}
}
\caption{\label{fig:res-all}Rank of synthetic document (averaged over all users) for all queries.}
\end{figure*}
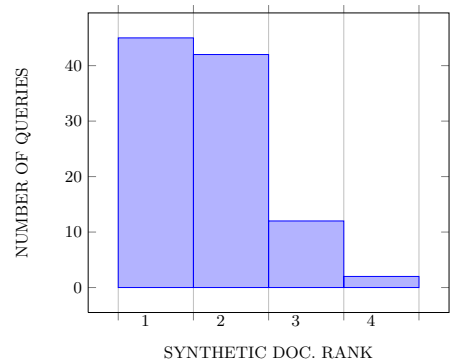
\begin{figure}
\centering
\scalebox{0.7}{
\begin{tikzpicture} \begin{axis}[ybar interval,
    x tick label style=
        {anchor=east},
    xlabel= SYNTHETIC DOC. RANK,
    ylabel = NUMBER OF QUERIES
    ]
\addplot coordinates
    {(1,45) (2,42) (3,12) (4,2) (5,0)};
\end{axis}
\end{tikzpicture}
}
\caption{\label{fig:bin}Average rank of synthetic documents (binned).}
\end{figure}

\begin{figure}
\centering
\scalebox{0.3}{
\includegraphics[]{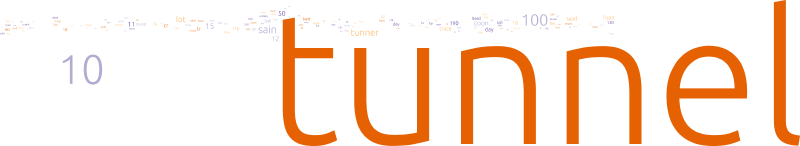}
}
\caption{\label{fig:bad1}Wordcloud of synthetic document for query: \textit{transportation tunnel disasters}.}
\centering
\scalebox{0.3}{
\includegraphics[]{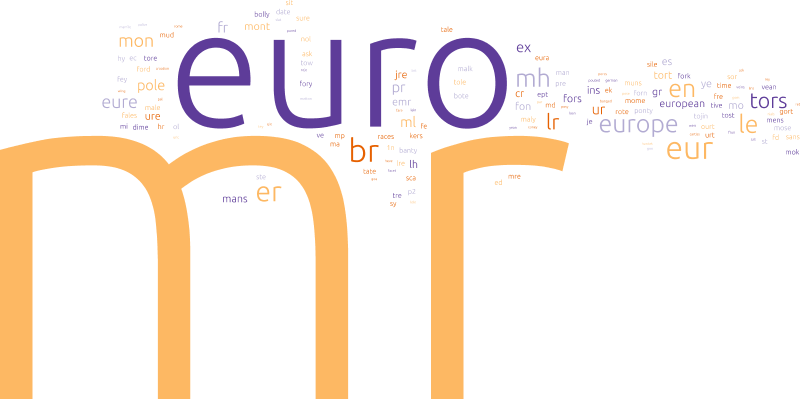}
}
\caption{\label{fig:bad2}Wordcloud of synthetic document for query: \textit{euro opposition}.}
\end{figure}

\subsection{Findings}

We report the average rank position of the synthetic document's wordcloud across all valid users for each query. This gives often a floating point rank position, e.g. rank position 1.5 in the example in Figure \ref{fig:example}. Figure \ref{fig:res-all} displays the average rank positions of all the synthetic document wordclouds, and Figure \ref{fig:bin} displays the same data but binned into rank positions 1, 2, 3 and 4. This binning aggregates positions 1.0 - 1.5 as 1, 1.6 - 2.5 as 2, 2.6 - 3.5 as 3, and 3.6 - 4.0 as 4.  We see that most synthetic document wordclouds are ranked at position 1 (for 45 out of 101 queries), followed closely by those ranked at position 2 (for 42 queries). The average rank position of all synthetic wordclouds for all queries is 1.81. This indicates that the wordclouds of the synthetic documents were assessed by users as most relevant of all displayed relevant documents on average.

Interestingly, there are only 2 synthetic documents whose wordclouds are ranked at position 4, and they are shown in Figures \ref{fig:bad1} \& \ref{fig:bad2}. For these two synthetic documents, one or two words had a significantly higher frequency than the remaining words, and as a consequence dominated the wordclouds, making almost all other words illegible. In both these cases, it makes sense that users would rank these wordclouds last, even if the dominating terms are relevant to the query (see the captions of Figures \ref{fig:bad1} \& \ref{fig:bad2}).


\section{Discussion}
\label{s:discussion}
Our idea of ``deep learning'' a single relevant document in response to a query may be reminiscent of question answering (QA) or summarisation-based IR systems. Unlike these approaches, our aim is to create a new document on some topic (not necessarily a question). Our approach differs in that this new document is synthesised by having learned the underlying semantics of the training documents, instead of extracting relevant parts from existing documents. 

Our findings show that the wordcouds of the synthetic documents were on average assessed more relevant than the wordclouds of existing relevant documents. We used wordclouds, as opposed to showing the full documents or document snippets, so that human assessors could have an easier and faster overview of the document contents. By doing so, we in fact limited the user understanding of the document contents (synthetic or not) to just the meaning of its 150 most frequent terms, without any term dependence or other co-occurrence information. This approach disregards deeper semantics.

The synthetic document of each query was ``deep learned'' by training on known relevant documents for that query. In the absence of such previously labelled data, relevant documents can be replaced by the top-$k$ retrieved documents for a query (similarly to how pseudo relevance feedback approximates relevance feedback), or even click-based approximations of relevant documents. We leave these options for future investigation. 

Finally, once a synthetic relevant document has been deep learned for a query, there are different ways of using it. In this work, we simply displayed it to the user (together with other known relevant documents) as word clouds, in order to assess its relevance. Another option is to use such a synthetic document to rank all the documents in the collection, by computing their semantic distance from it and ranking them accordingly. We experimented with this idea, but found it ineffective for this setup. We found it much more effective for reranking top-retrieved documents, than for ranking all documents in the index, similarly to \cite{MitraNCC16}.

\section{Conclusion}
\label{s:conc}
We proposed an Information Retrieval (IR) setup, where the IR system does not retrieve existing information, but instead learns from it and produces a single relevant document that, we assume, encompasses all its indexed relevant information for a query. 
While a decade ago such a setup might have sounded science-fictional, the advances of deep learning have made this possible. Specifically, given a query, we trained a character-level Long Short Term Memory (LSTM) Recurrent Neural Network (RNN) on all its indexed relevant documents, and used it to output a new, synthetic document. We then showed the query and wordcloud representations of three randomly chosen existing relevant documents and of our new synthetic document to users, and asked them to rank them by relevance. The synthetic document was overall ranked highest across all queries and users. 

In the future, we plan to experiment with word-level RNNs, which are expected to produce even better quality synthetic documents (character-level RNNs have been criticised for outputting noisy pseudo-terms \cite{BojanowskiJM15}).

\paragraph{Acknowledgements} This work was supported by the Danish Industry Foundation and the Industrial Data Analysis Service (IDAS) project.
\bibliographystyle{abbrv}
\bibliography{dl}  
%
%
\end{document}